\documentclass[useAMS]{mn2e}

 \usepackage[usenames,dvipsnames]{color}

\usepackage{graphicx}

\title[Variability in the CSPN NGC2392]{Rummaging inside the Eskimo's parka:
Variable asymmetric PN fast wind and a binary nucleus?
}
\author[R.K. Prinja {\&} M. Urbaneja]{R. K. Prinja$^{1}$, M. A. Urbaneja$^{2,3}$,\\
$^{1}$Dept. of Physics {\&} Astronomy, University College London, Gower Street, London WC1E 6BT \\
$^{2}$Institute for Astro- and Particle Physics,
University of Innsbruck,
Technikerstr. 25/8, A-6020 Inssbruck, Austria \\
$^{3}$Institute for Astronomy, University of Hawaii, 2680 Woodlawn Drive, 
Honolulu, HI 96822, USA \\
}
\begin{document}

\date{Accepted 2009. Received 2009; in original form 2009}

\pagerange{\pageref{firstpage}--\pageref{lastpage}} \pubyear{2007}

\maketitle

\label{firstpage}

\begin{abstract}
We report on high-resolution optical time-series spectroscopy
of the central star of the `Eskimo' planetary nebula NGC~2392.
Datasets were secured with the ESO 2.3m in 2006 March and
CFHT 3.6m in 2010 March to diagnose the fast wind and photospheric
properties of the central star. The He{\sc i} and He{\sc ii}
recombination lines reveal evidence for clumping and temporal structures
in the fast wind that are erratically variable on timescales
down to $\sim$ 30 min. (i.e. comparable to the characteristic
wind flow time).
We highlight changes in the overall morphology of the wind lines
that cannot plausibly be explained by line-synthesis model
predictions with a spherically homogeneous wind.
Additionally we present evidence that the UV line profile morphologies
support the notion of a high-speed, high-ionization polar wind
in NGC~2392.
Analyses of deep-seated, near-photospheric absorption lines
reveals evidence for low-amplitude radial velocity shifts. 
Fourier analysis points tentatively to a 
$\sim$ 0.12-d modulation
in the radial velocities, independently evident in the ESO and CFHT data.
We conclude that the overall spectroscopic properties
support the notion of a (high inclination) binary nucleus in
NGC~2392 and an asymmetric fast wind.
\end{abstract}

\begin{keywords}
stars: mass-loss $-$ stars: evolution $-$ stars: individual: NGC~2392
$-$ optical: stars
\end{keywords}

\section{Introduction}

Planetary nebulae (PNe) are a key pathway in the evolution of low to
intermediate mass stars, and their central stars are the immediate
precursors of white dwarfs. Studies of PN central stars (herein CSPNe)
are motivated by: The desire to understand the origin of the rich
variety of PN morphologies; to establish the mass-loss process via
fast winds driven by radiation pressure by spectral lines; and
to secure fundamental stellar parameters that can test post-AGB stellar
evolution models.

Time-series spectroscopy is an important diagnostic tool in developing
our understanding of CSPNe. Recently, far-UV and UV datasets have revealed
signatures of large-scale wind structures and evidence for modulated
temporal behaviour that may provide a handle on the central star
rotation rates (e.g. Prinja et al. 2012a, 2012b). Similarities between the
wind properties of H-rich CSPNe and those of massive Population I OB stars
(which also have line-driven winds) suggest that instabilities in variable
fast winds may result in shock heated gas which emits X-rays in the central
cavities of PNe (e.g. Guerrero 2006; Kastner et al. 2012).
In the optical waveband, time-series data are requisite for establishing
systematic radial velocity shifts in CSPNe absorption lines. De Marco et al. (2004)
have for example conducted a radial velocity survey of 11 CSPNe to provide
constraints on the binary properties of the parent AGB population and thus
the extent to which binarity may play a causal role in shaping non-spherical
nebulae.

In this paper we present time-series optical spectra of the central star 
of NGC~2392 (Eskimo nebula). Our study is motivated by several interesting 
characteristics, discrepancies and scenarios for this PN: (i) The central 
star of NGC~2392 exhibits high He, N and low C, O abundances suggesting 
that the photosphere has been processed (M{\'e}ndez et al. 2012). A 
possible scenario is that the abnormal central star abundances are due to 
a common-envelope evolutionary phase thus implying a close binary 
companion; (ii) Danehkar et al. (2011) employ photoionization models of 
high excitation PN emission lines to argue that NGC~2392 has a hot white 
dwarf ($\sim$ 1 M$_\odot$) companion; (iii) Detailed kinematic modelling 
of the (Eskimo) nebula by Garcia-Diaz et al. (2012) supports a near-pole 
orientation, complex nebula morphology with multiple kinematic
components, and an evolution path that may invoke a common-envelope 
binary; (iv) The extended and point X-ray emission from NGC~2392 (e.g. 
Kastner et al. 2012) is not entirely consistent with the predicted thermal 
energy converted from the kinetic energy of the fast wind. Additional 
coronal energy from a binary companion may explain the observed high X-ray 
temperatures.

Despite all the implications of the above studies, there is no definitive
evidence so far of a binary nucleus in NGC~2392, and the time-variable
and geometric characteristics of its fast wind are not established.
In this study we present the analysis of high-resolution optical
time-series datasets secured over two epochs in 2006 and 2010
using the 3.6m ESO and Canada-France-Hawaii (CFHT) observatories.
Our goal is to investigate for the first time $\sim$ hourly changes
in the fast wind of NGC~2392 and fluctuations close to the surface of
the central star. We characterise here evidence for evolving
structure in the outflow and indications of radial velocity changes
in deep-seated absorption lines.

\section{Optical spectroscopy}

A log of the time-series spectra of the central star of NGC~2392 is
given in Table 1. Thirteen spectra were secured over 3 consecutive nights
in 2010 March using the ESPaDonS echelle spectrograph
(Manset {\&} Donati, 2003; Donati et al. 1997)
on the 3.6m CFHT
at Mauna Kea, Hawaii.
The continuum signal-to-noise ratio (S/N) of an individual
spectrum is $\sim$ 100 for 30 min exposures, with a
spectral resolution, R, $\sim$ 68000.
The data were reduced using the standard CFHT pipeline Upena.
The CFHT data are complemented in this study by 18 spectra obtained at
the ESO La Silla 3.6m telescope using the HARPS echelle spectrograph
(Mayor et al., 2003).
The ESO observations were carried out by us during 2006 March
over 3 consecutive nights (Table 1). Typical individual HARPS
spectra have S/N $\sim$ 40 (for 30 min exposures) and R $\sim$ 110,000.
The ESO automatic online pipeline was used for homogeneous reduction.


\begin{table}
 \centering
\caption{Log of observations.}
  \begin{tabular}{lll}
  \hline
Observatory & MJD range (days) & No. of spectra  \\
\hline

ESO HARPS & 53818.002 $-$ 53818.109 & 6 \\
          & 53818.995 $-$ 53819.101 & 6 \\
          & 53820.002 $-$ 53820.108 & 6 \\
\\
CFHT ESPaDOnS & 55256.358 $-$ 55256.447 & 5 \\
              & 55257.271 $-$ 55257.337 & 4 \\
              & 55258.270 $-$ 55258.336 & 4 \\
\hline

\end{tabular}
\end{table}
  

All the spectra were normalised by fitting a low-order polynomial
through continuum windows. The line profiles discussed here have been
corrected for a radial velocity of 82 km s$^{-1}$, measured in
weak absorption lines. The fundamental central star
parameters adopted in this study are listed in Table 2.


\begin{table}
 \centering
\caption{NGC~2392 adopted central star parameters.}
  \begin{tabular}{lll}
  \hline
Parameter & Value & Reference  \\
\hline

Sp. type & Of (H-rich) & M{\'e}ndez et al. (1991) \\
$v_\infty$ & 400 km s$^{-1}$ & Kaschinski et al. (2012) \\
$v$sin($i$) & 50 km s$^{-1}$ & M{\'e}ndez et al. (2012) \\
T$_{\rm eff}$ & 43000 K  & M{\'e}ndez et al. (2012) \\
R$_\star$ & 1.9 R$_\odot$ & M{\'e}ndez et al. (2012)  \\
Radial velocity & 82 km s$^{-1}$ & This paper \\
\hline
\end{tabular}

$v_\infty$ =  wind terminal velocity; $v$sin($i$) = projected rotation velocity.

\end{table}
  

\begin{figure*}
 \includegraphics[scale=0.33]{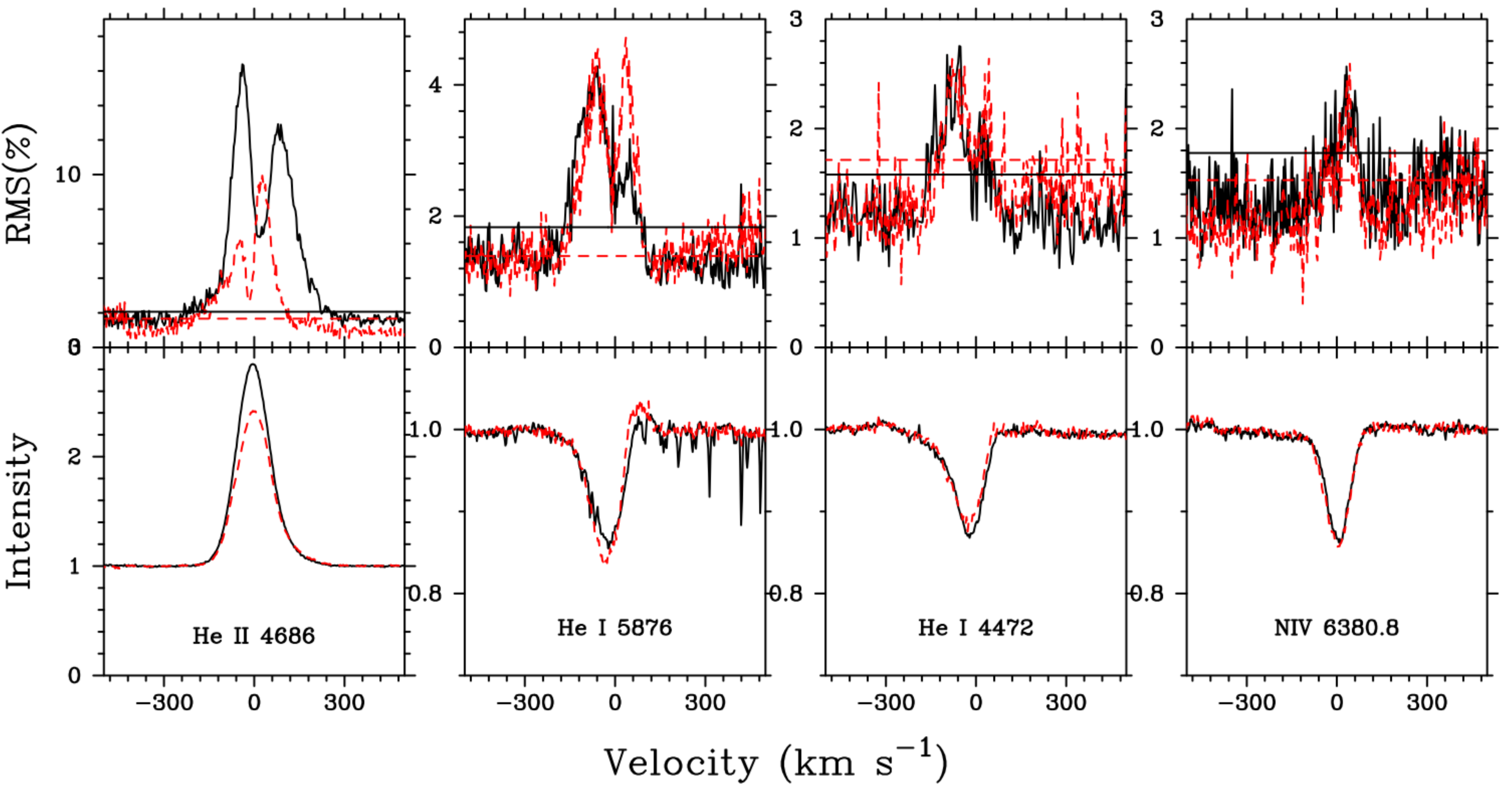}

 \caption{Variability in the
He{\sc ii} $\lambda$4686, 
He{\sc i} $\lambda$5876, He{\sc i} $\lambda$4472 and
N{\sc iv} $\lambda$6381 line profiles of NGC~2392.
For each case, the upper panel displays the
total variance spectrum (TVS) and 95{\%}
confidence limit (horizontal lines). The lower
panels below
the TVS show the mean line profiles
for the ESO (solid black line)
and CFHT (red dashed lines) time-series data.
(The same colour and line scheme is also
used in the TVS panels.)
}

\end{figure*}

\section{Time-variable fast wind}

As is generally the case with hot stars, radiation pressure driven winds are
ideally diagnosed via UV resonance lines formed by the re-distribution
of the stellar continuum radiation through line scattering.
High-resolution {\it International Ultraviolet Explorer} ($IUE$) and
{\it Far-Ultraviolet Spectroscopic Explorer} ($FUSE$) satellite data
of the central star in NGC~2392 have been presented by
e.g. Patriarchi {\&} Perinotto (1995); Herald {\&} Bianchi (2011)
and Kaschinski et al. (2012). The consensus is that the fast wind
has a terminal velocity, $v_\infty$, $\sim$ 400 km s$^{-1}$ and
(smooth wind) mass-loss rate $\sim$ 3$-$5 $\times$ 10$^{-8}$ M$_\odot$ yr$^{-1}$.

\begin{figure*}
\begin{tabular}{cc}
\vspace*{-2in} 
\hspace*{-0.6in}
\includegraphics[scale=0.45]{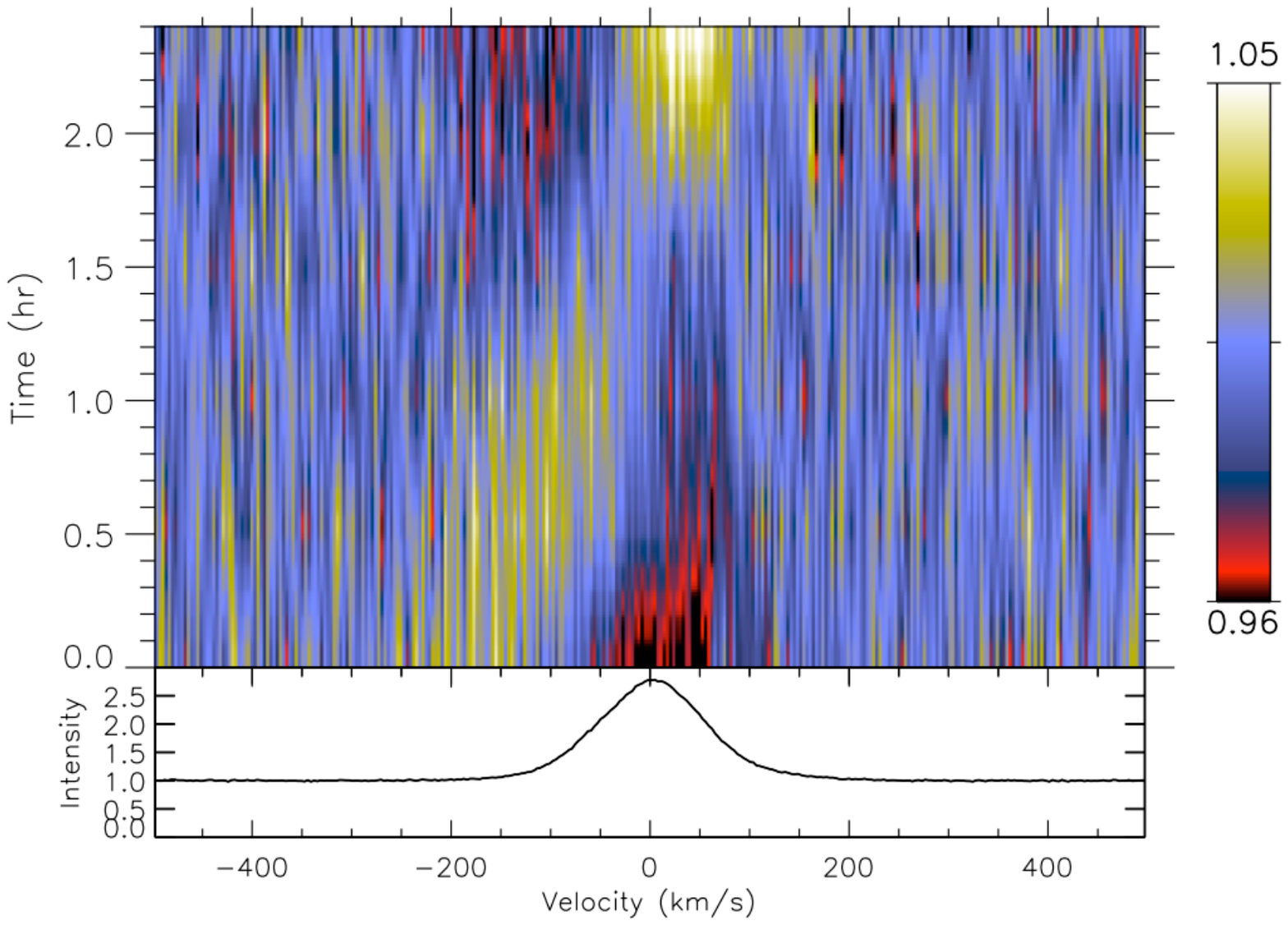} &
\includegraphics[scale=0.45]{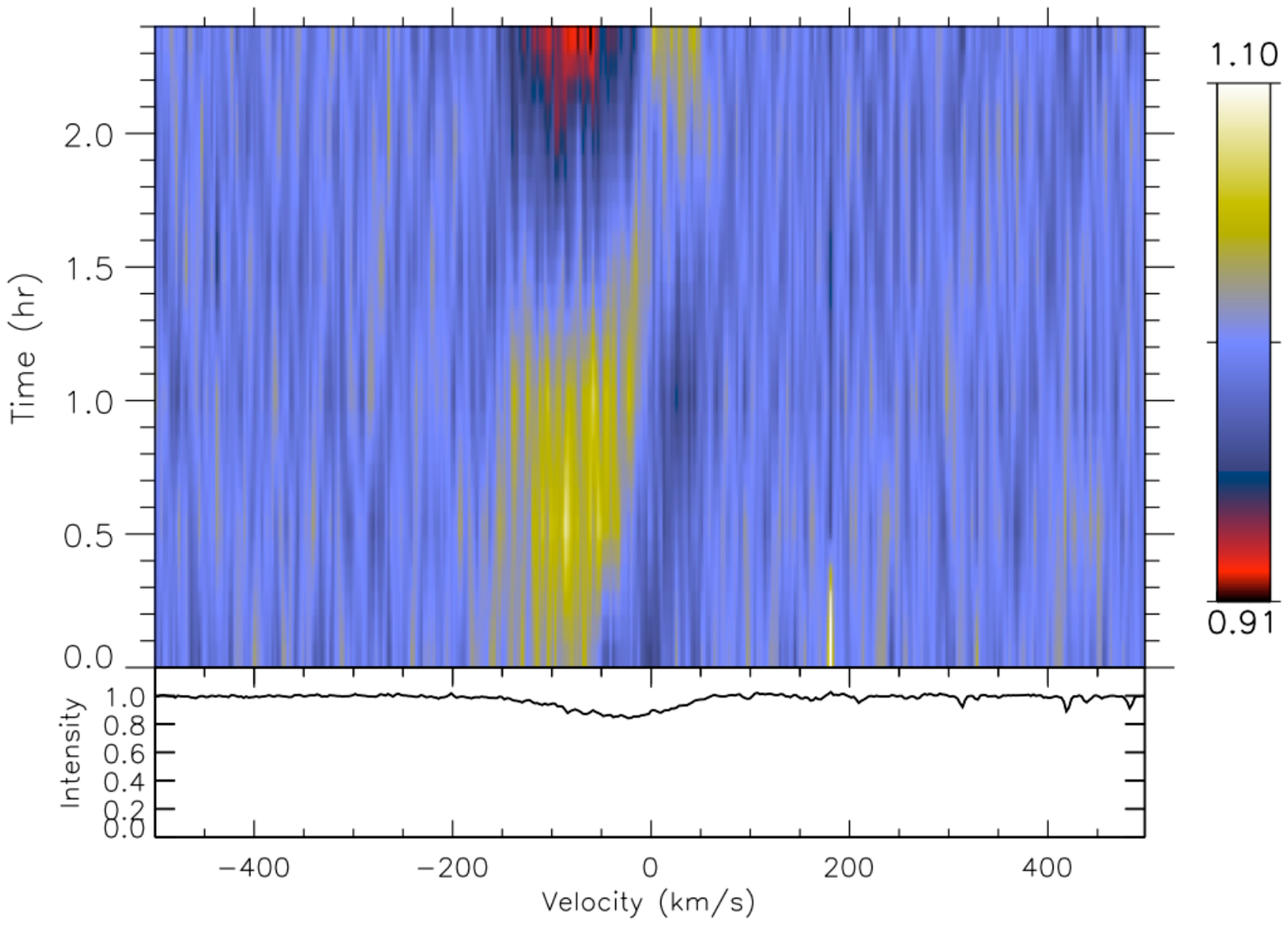} \\
\vspace*{-2in}
\hspace*{-0.6in}
 \includegraphics[scale=0.45]{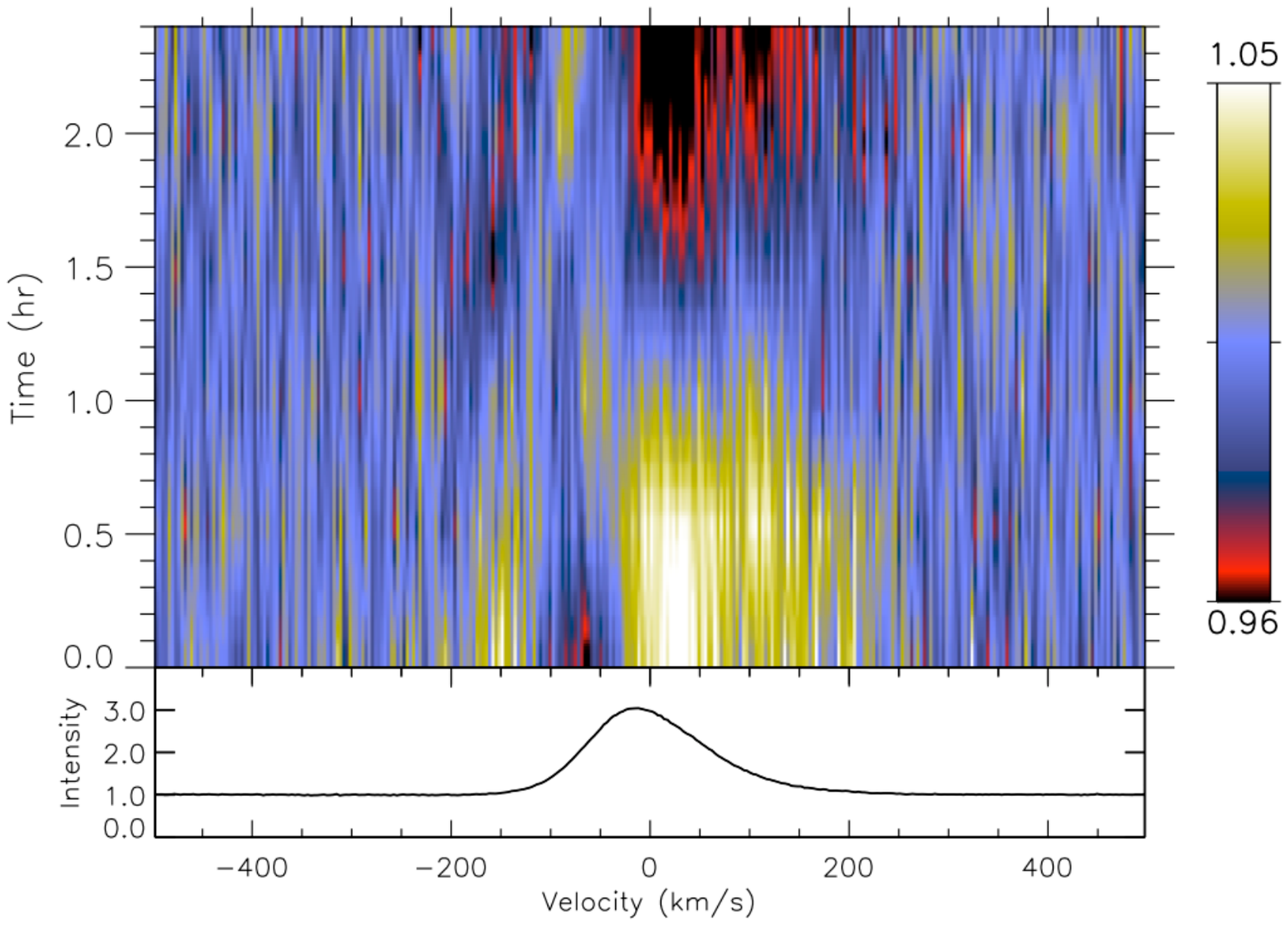} &
\includegraphics[scale=0.45]{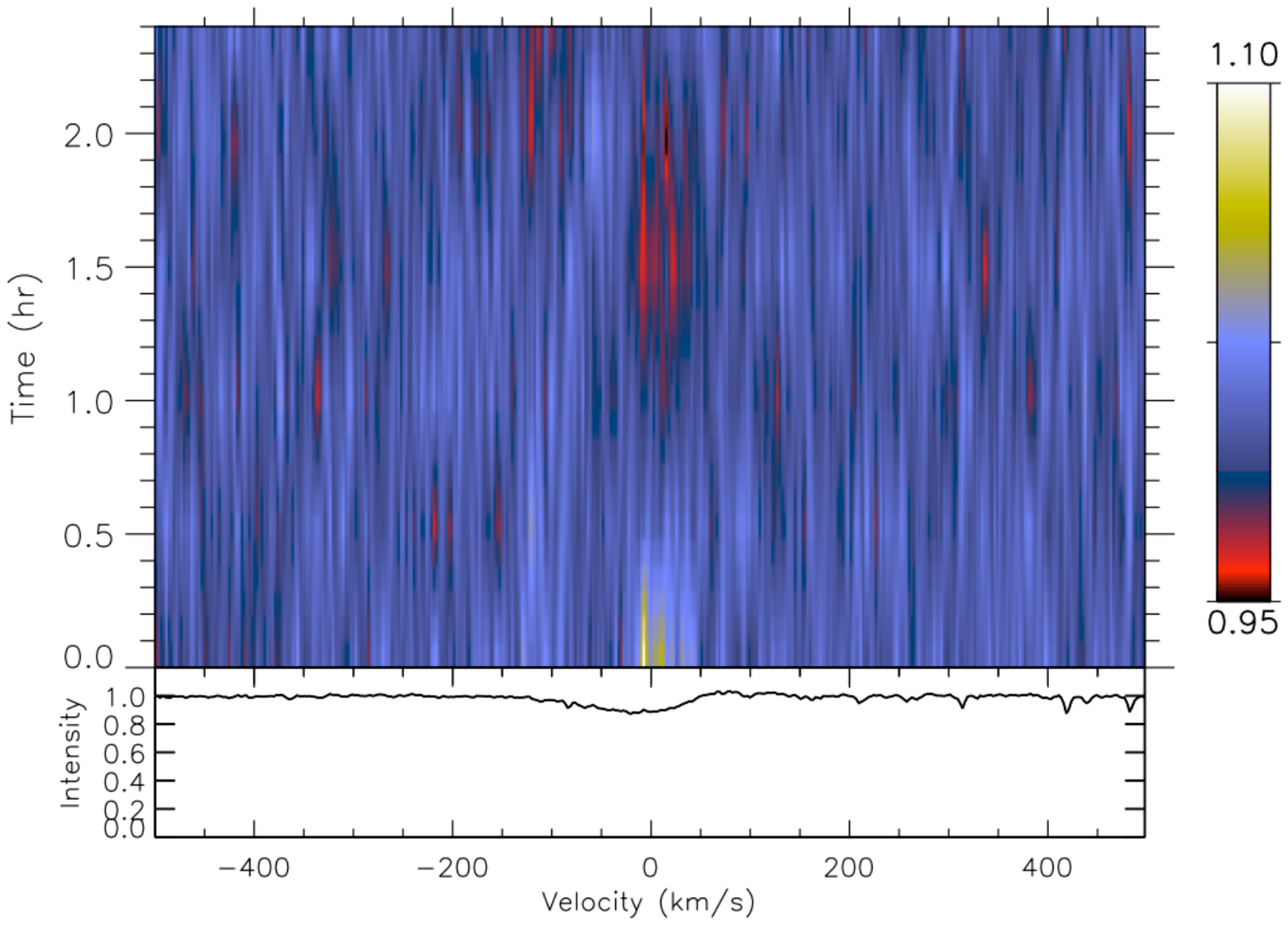} \\
\vspace*{-2in}
\hspace*{-0.6in}
  \includegraphics[scale=0.45]{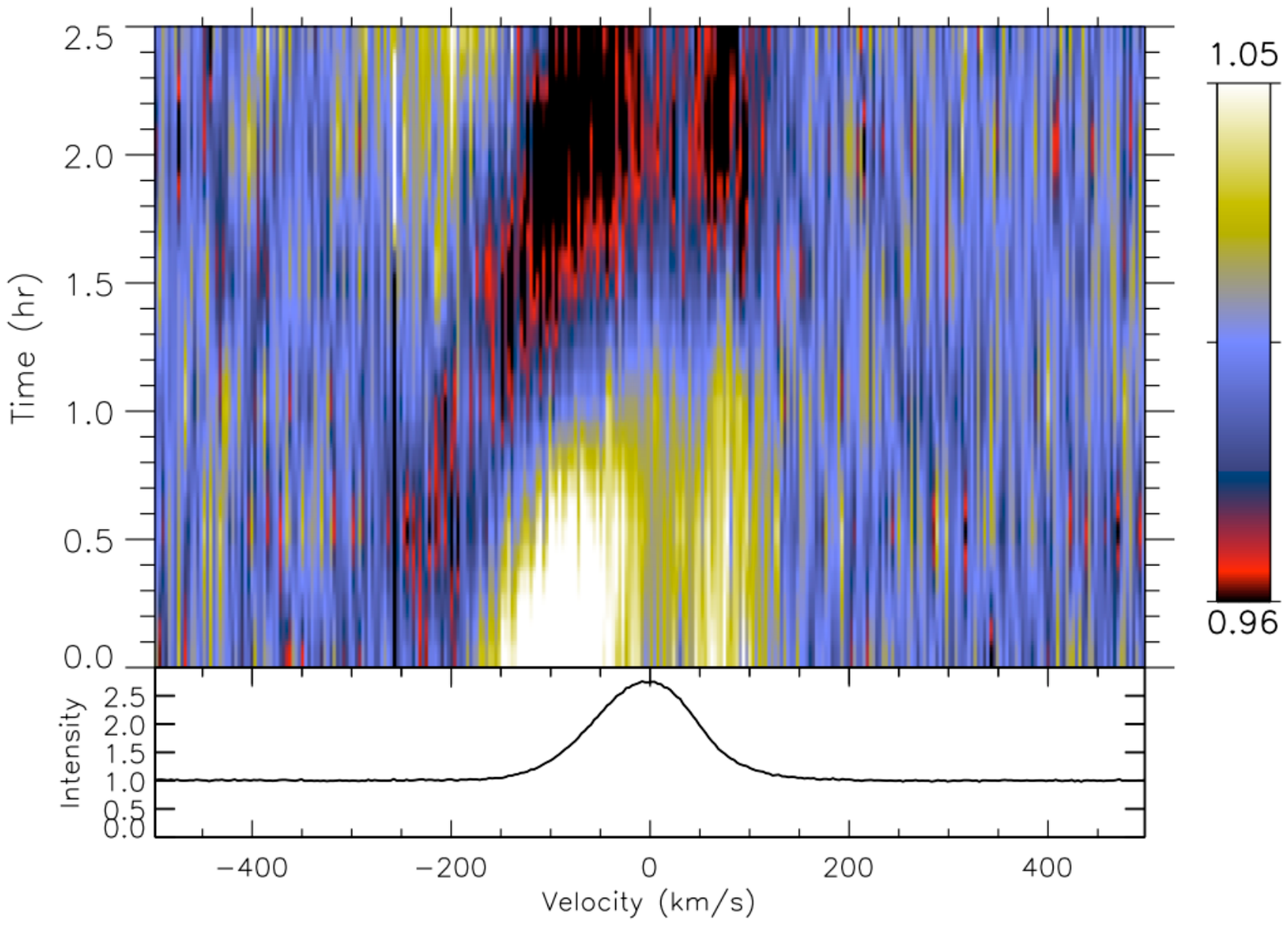} &
\includegraphics[scale=0.45]{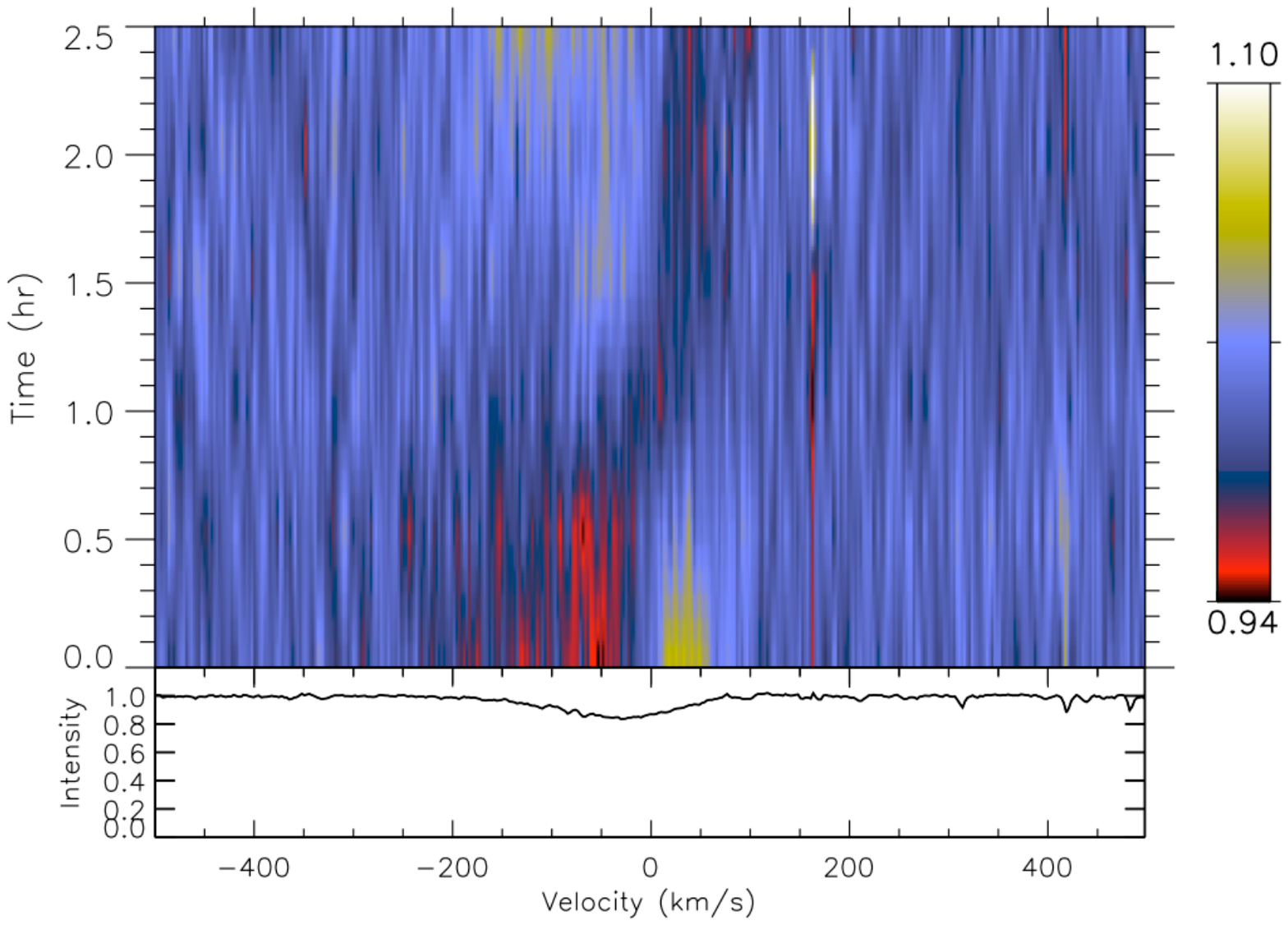} \\
\end{tabular}
 \caption{Dynamic spectra showing the nightly line profile
variability in the ESO (2006) data of NGC~2392.
The left-hand and right-hand panels are
for He{\sc ii} $\lambda$4686 and He{\sc i} $\lambda$5876, respectively.
}
\end{figure*}

\begin{figure*}
\begin{tabular}{cc}
\vspace*{-2in} 
\hspace*{-0.6in}
\includegraphics[scale=0.45]{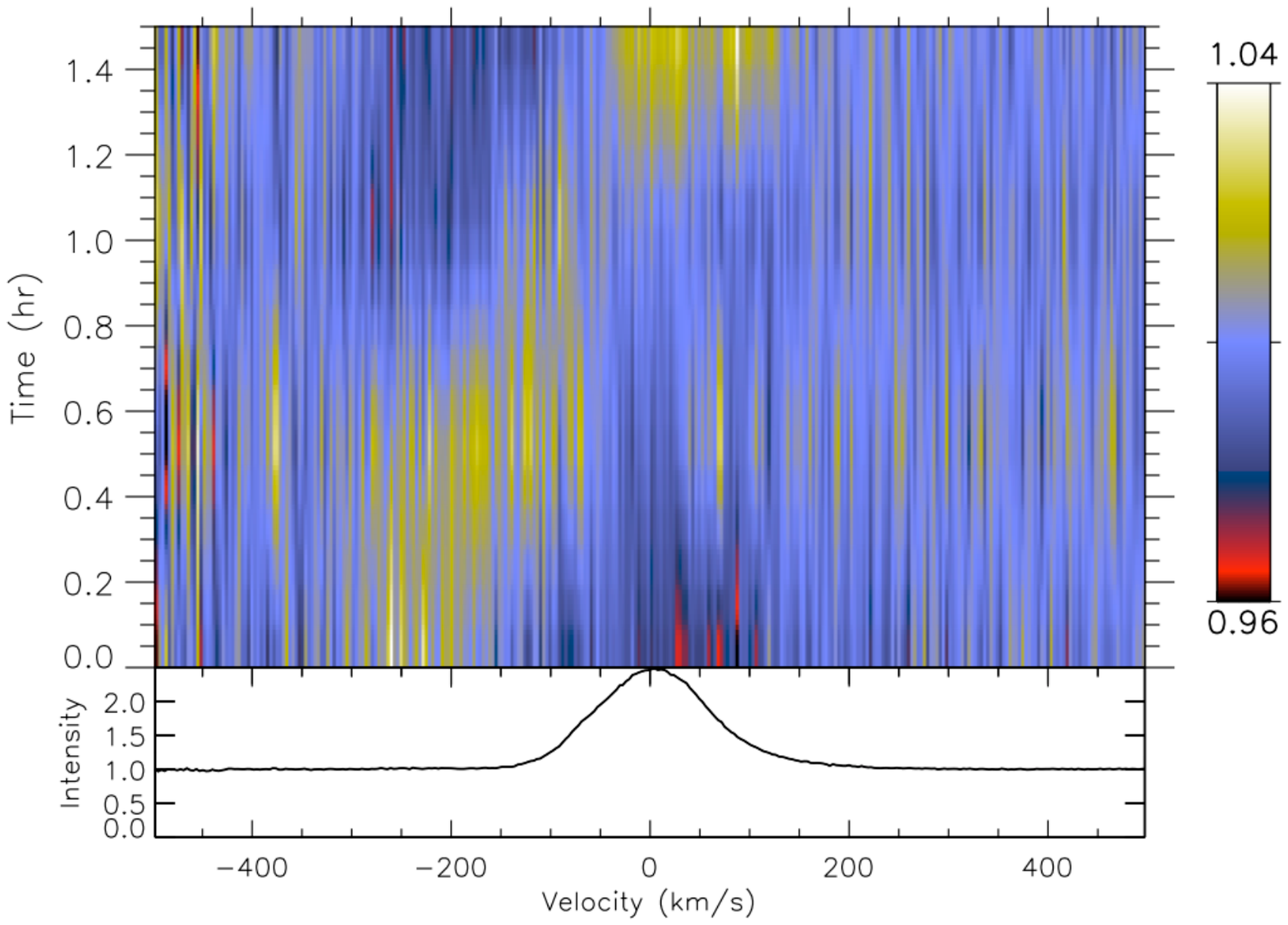} &
\includegraphics[scale=0.45]{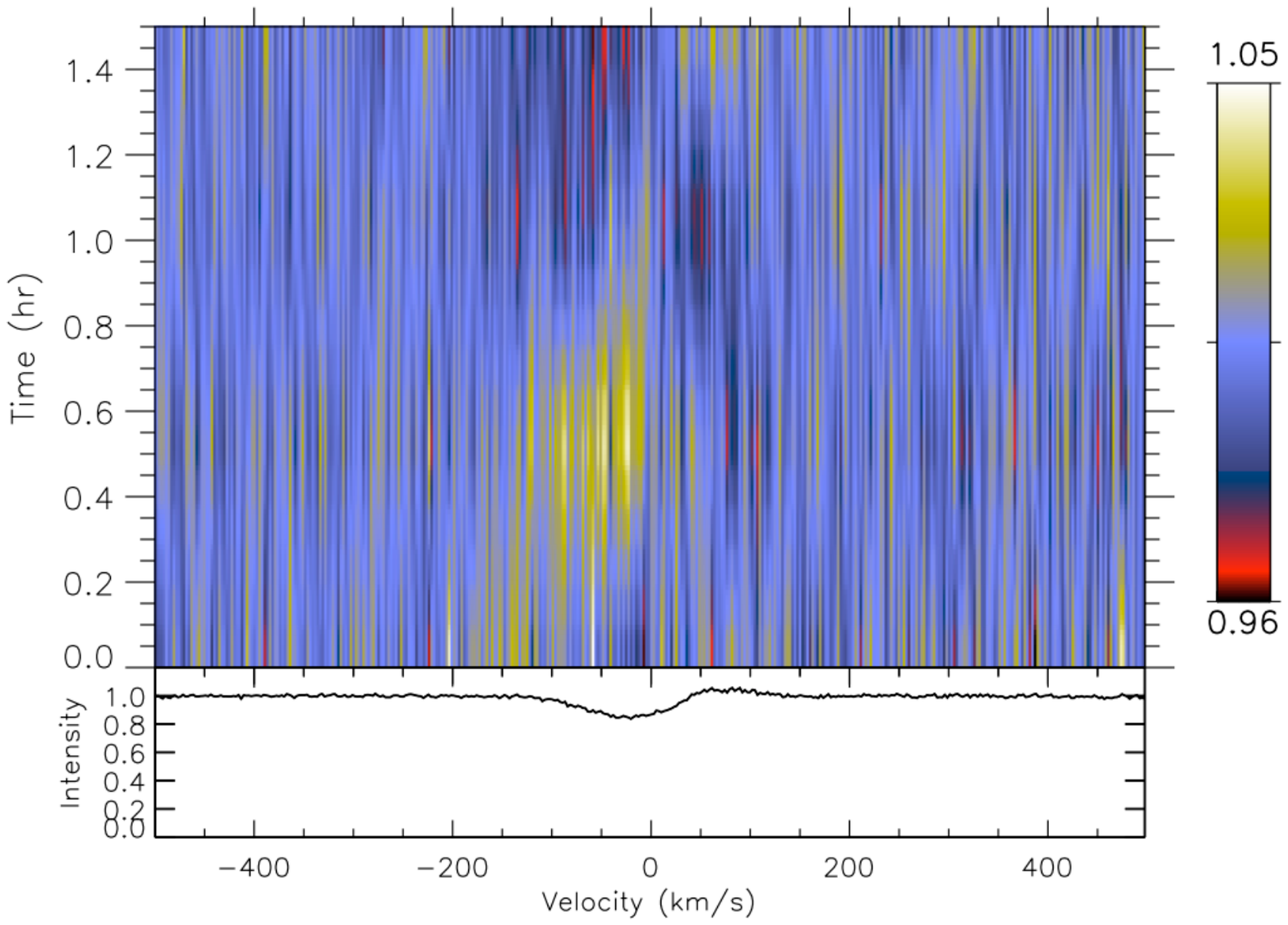} \\
\vspace*{-2in}
\hspace*{-0.6in}
 \includegraphics[scale=0.45]{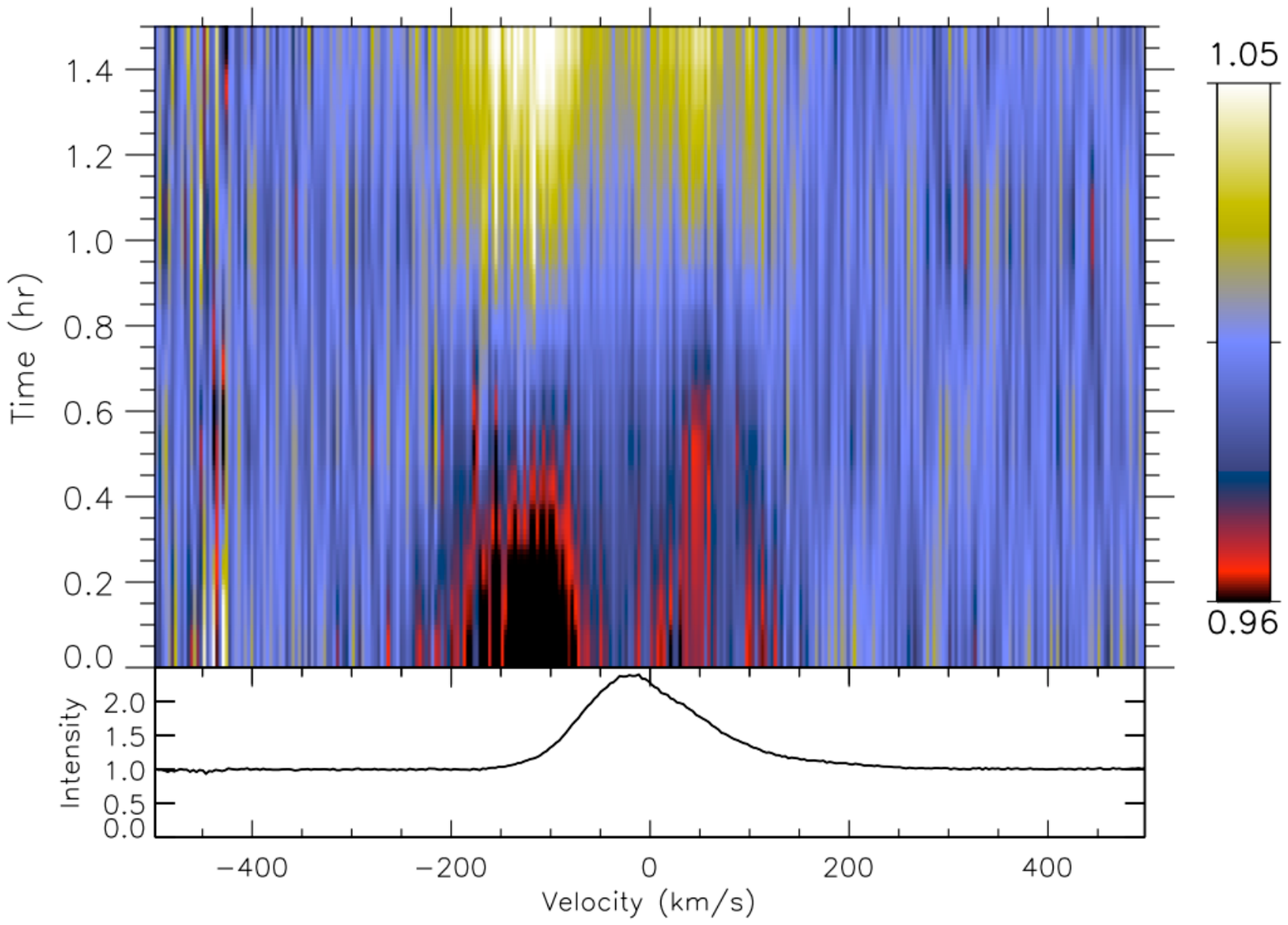} &
\includegraphics[scale=0.45]{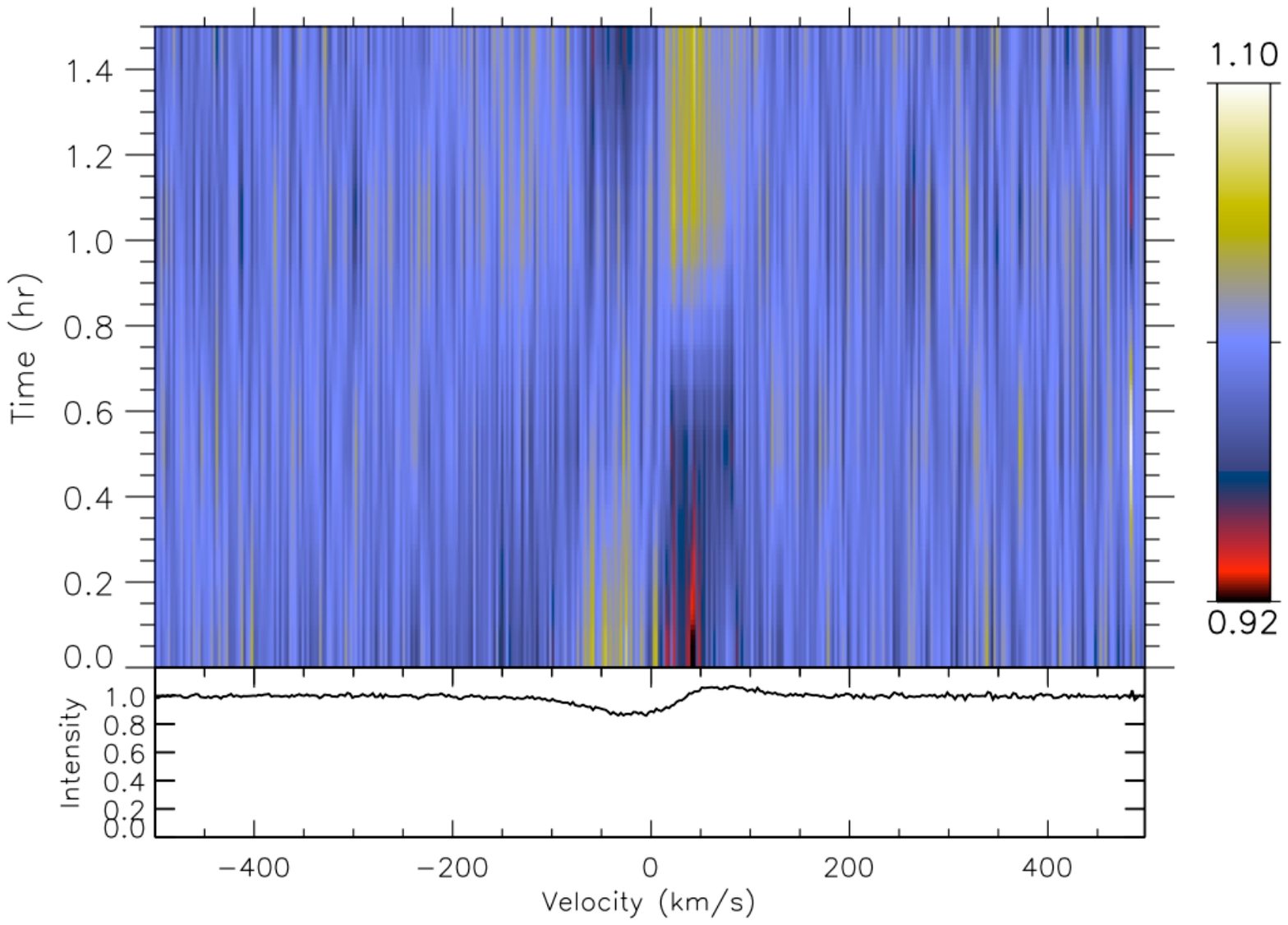} \\
\vspace*{-2in}
\hspace*{-0.6in}
  \includegraphics[scale=0.45]{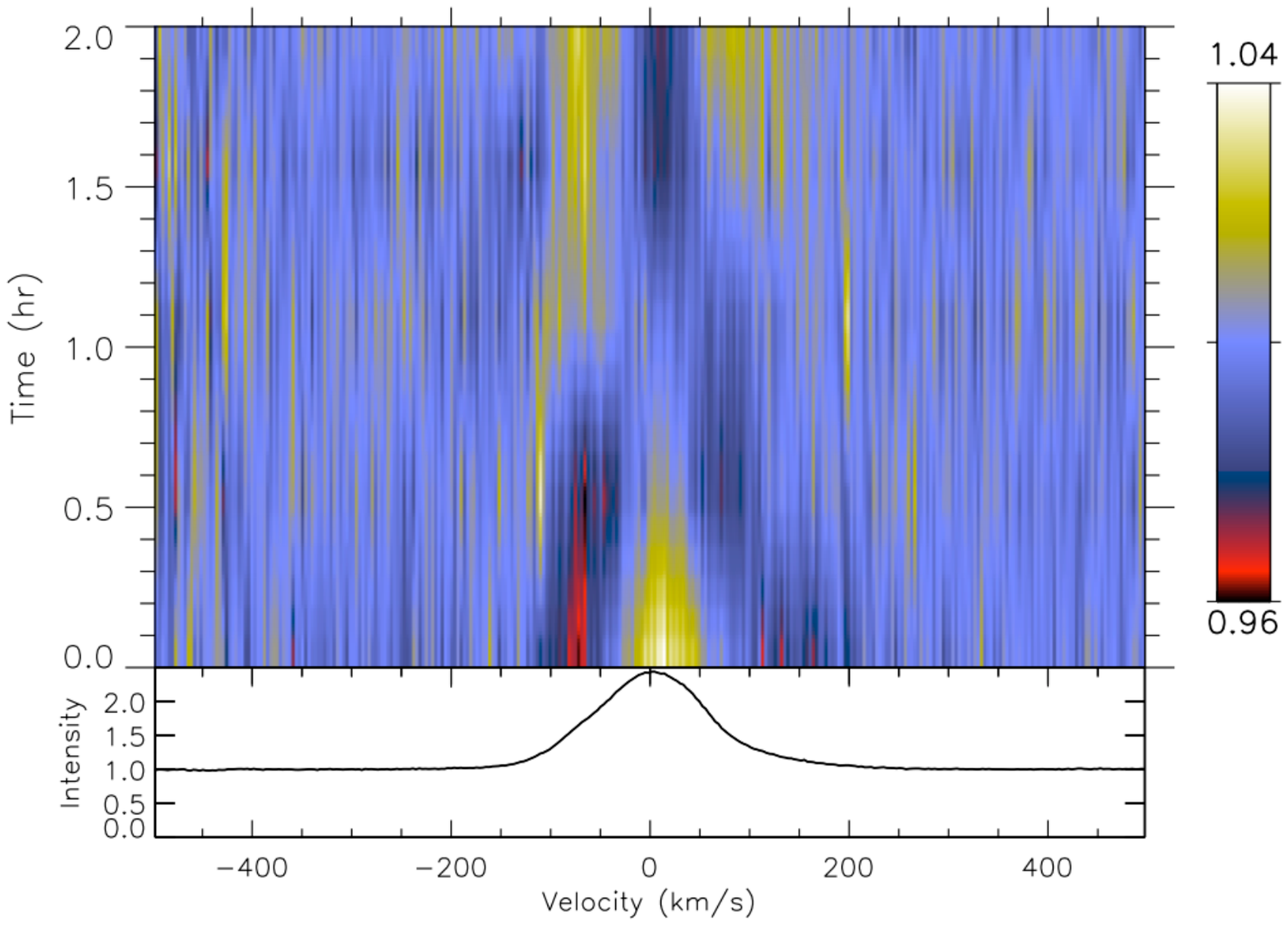} &
\includegraphics[scale=0.45]{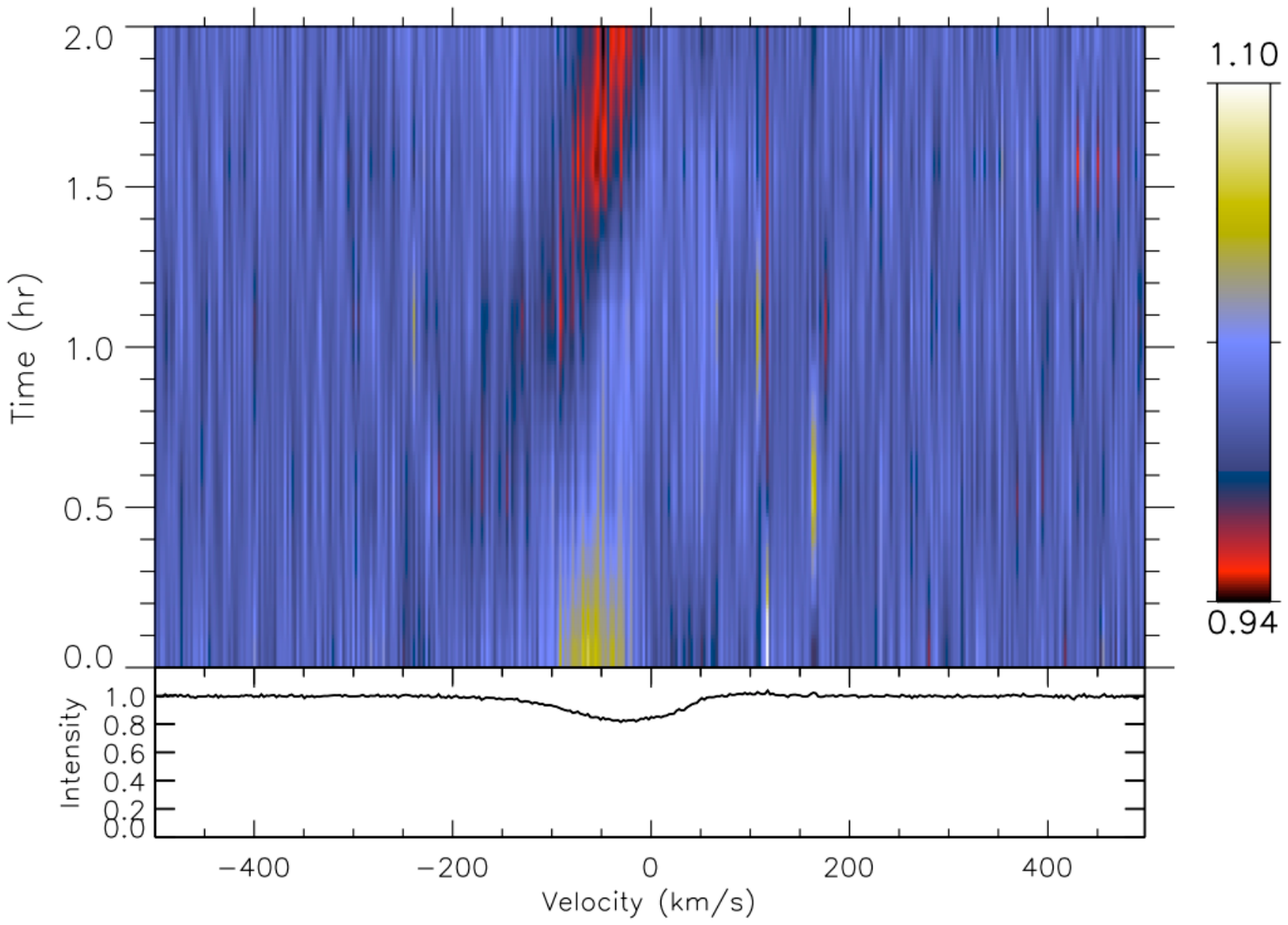} \\
\end{tabular}
 \caption{Dynamic spectra showing the nightly line profile
variability in the CFHT (2010) data of NGC~2392.
The left-hand and right-hand panels are
for He{\sc ii} $\lambda$4686 and He{\sc i} $\lambda$5876, respectively.
}
\end{figure*}

There are no archival FUV or UV {\it time-series} datasets available
for NGC~2392 that permit an investigation of systematic variability
on time-scales comparable to the wind flow time (i.e. $\sim$ hours).
Patriarchi {\&} Perinotto (1995) show sparse $IUE$ spectra separated
by $\sim$ 7 years which suggest some fluctuations in the
C{\sc iv} and N{\sc v} lines, and Guerrero {\&} De Marco (2013)
comment on the two {\it FUSE} spectra available for the central star.

The optical time-series data of the nuclei of NGC~2392 that we
present here provide access to the fast wind via excited transitions
arising from He{\sc i} and He{\sc ii}. 
(Note that the H$\alpha$ and H$\beta$ lines are
heavily affected by nebular emission.)
The He lines are primarily the
result of recombination from a higher ionization stage followed by
a radiative de-excitation. Since recombination is sensitive to
the square of wind density ($\rho^2$), the optical lines form in the
densest region of the wind (in contrast to the UV lines).
Furthermore the He{\sc i} $\lambda$5876 (2$^3$P$^0$ $-$ 3$^3$D) transition
is a potential exception if large populations are acquired
in its lower level such that it effectively becomes the ground level
and consequently behaves as a scattering or resonance line.

We find that the stellar lines in the optical spectra of NGC~2392
are undoubtedly variable within $\sim$ hours, with peak-to-peak
amplitude changes of $\sim$ 5{\%} in He{\sc i} $\lambda$5876 and
He{\sc ii} $\lambda$4686.
The corresponding equivalent width and standard deviations for the
combined ESO and CFHT datasets are $\sim$ 0.32{\AA} s.d. 0.12{\AA}
and $\sim$ $-$3.8{\AA} s.d. 0.6{\AA}. The line profile variability
is quantified in Fig. 1 where we employ the
Temporal Variance Spectrum (TVS) method (e.g. Fullerton et al. 1996)
to estimate the statistical significance of the changes, having
accounted for differences in the data quality at spectrum and
pixel levels.
The variability level is higher in He{\sc ii} $\lambda$4686 than
He{\sc i} $\lambda$5876 or He{\sc i} $\lambda$4472; note that numerous
other stellar lines are also weakly variable in the night-to-night data.
Significant variability extends to $\sim$ $-$180 km s$^{-1}$ (i.e.
$\sim$ 0.45 $v_\infty$) in the ESO (2006) and CFHT (2010) spectra
of He{\sc ii} $\lambda$4686, but the redward extend of these changes is
lower in 2010 compared to 2006. For both epochs the
He{\sc i} $\lambda$5876 weak P~Cygni profile varies between
$\sim$ $-$150 km s$^{-1}$ to 100 km s$^{-1}$.
An important feature of the TVS in Fig. 1 is that it is {\it double-peaked}
in the He{\sc ii} and He{\sc i} lines for the ESO and CFHT datasets.
Typically the peaks in the TVS are separated by $\sim$ 100 to 120 km s$^{-1}$.
The amplitude of the TVS peaks are not systematically stronger or weaker
in either the blue or red sides. For line profile variability dominated
by clumps we may expect small-scale structure to be distributed evenly
over the whole inner wind thus presenting a broadly symmetrical single-peak TVS.
One possibility for the double-peak TVS is that it results from radial velocity
shifts perhaps due to a binary nature in NGC~2392. We return to the issue
of radial velocity changes in Sect. 4.

The systematics of the He{\sc ii} $\lambda$4686
and He{\sc i} $\lambda$5876 profile changes are examined
further in Figs. 2 and 3 where the data are displayed in
two-dimensional velocity$-$time dynamic spectra (images). The
intensity levels of individual spectra are determined as
{\it differences} with respect to the average spectrum for
each observing night, and represented by a colour scale
from minimum (black) to maximum (white) intensity cut level.
The features seen in Figs. 2 and 3 therefore represent
pseudo-absorption and pseudo-emission features relative to
the mean. The fast wind in NGC~2392 is undoubtedly variable on
very short time-scales and can switch between an overall
pseudo-emission to pseudo-absorption pattern in the dynamic spectra
in less than 30 minutes (i.e. time lengths comparable to
the characteristic wind flow time, $\sim$ $R_\star/v_\infty$).

The evidence from Figs. 2 and 3 is that episodes of strong
blue and red changes in He{\sc ii} $\lambda$4686 are
mimicked in He{\sc i} $\lambda$5876 at the same velocities.
For example enhanced redward emission in He{\sc ii} $\lambda$4686 is
accompanied by an increase in the (weak) P~Cygni emission component
in He{\sc i} $\lambda$5876. A substantial low velocity blueward
emission increase in He{\sc ii} $\lambda$4686 is `matched' by
an increase in the P~Cygni absorption strength in He{\sc i} $\lambda$5876.
The overall impression from our optical time-series is that the fast wind in
NGC~2392 is stochastically variable over $\sim$ hours as opposed
to revealing coherent modulated behaviour. We do not for example see
signs of blueward and/or redward migrating features during each night,
nor is there evidence for much slower evolving (large-scale) structures 
in the wind that persist over $\sim$ 3 nights.

\begin{figure}
 \includegraphics[scale=0.37]{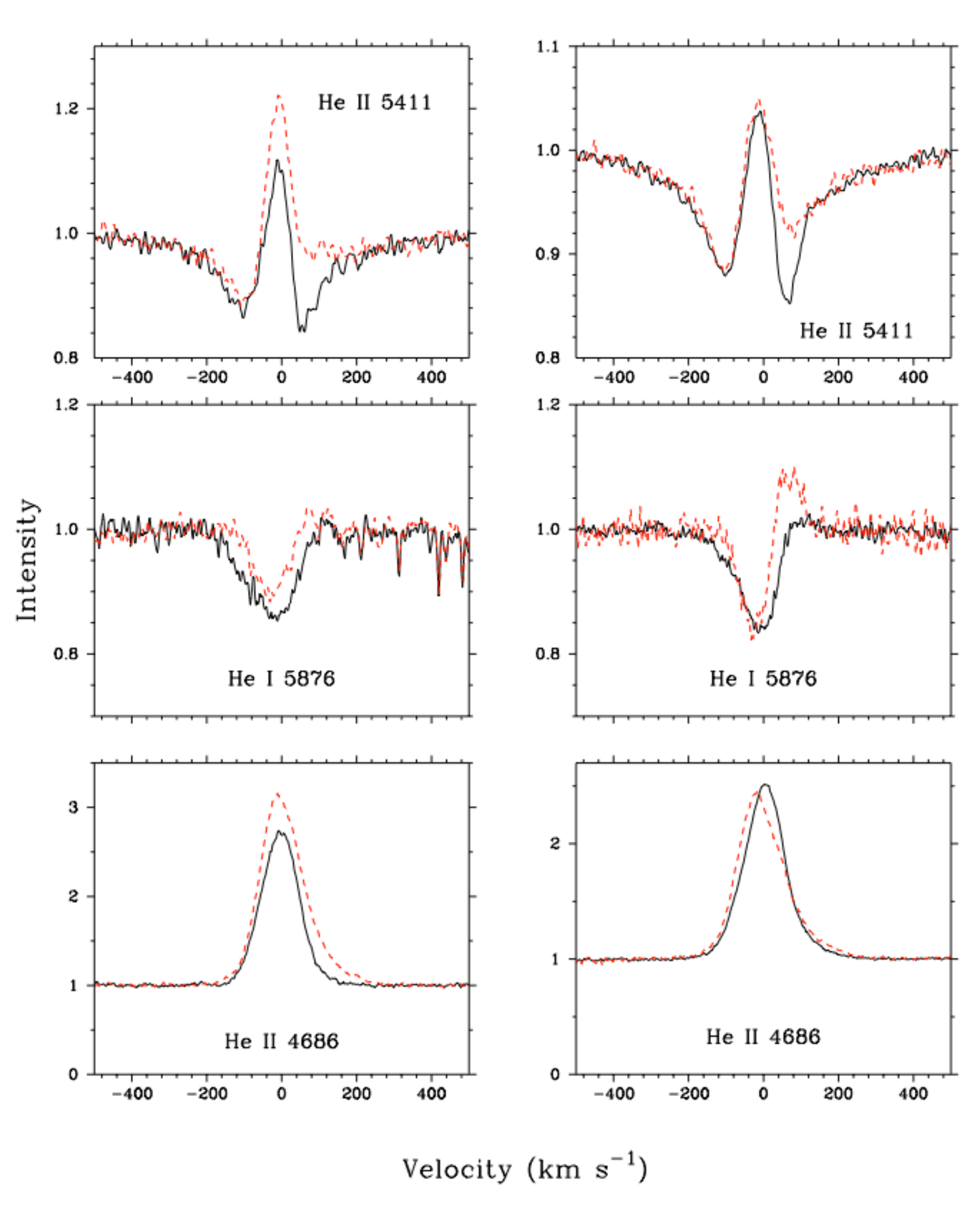}

 \caption{The {\it corresponding} changes in line
profile morphology evident between pairs of
He{\sc ii} $\lambda$5411, He{\sc i} $\lambda$5876 and He{\sc ii} $\lambda$4686
line profiles taken $\sim$ 1 day apart.
The left- and right-hand panels show
examples from the ESO
and CFHT data, respectively.
}

\end{figure}

The overall line profile morphology can change in complex manners from
night-to-night. We show in Fig. 4 corresponding pairs of He{\sc i} $\lambda$5876,
He{\sc ii} $\lambda$5411
and He{\sc ii} $\lambda$4686 spectra. Each pair is separated by $\sim$ 1 day.
The left-hand panel in Fig. 4 shows a case where as the
total He{\sc i} $\lambda$5876 absorption increases across all velocities,
in the corresponding He{\sc ii} $\lambda$4686 pair the overall emission
decreases. For this same pair, the changes in
He{\sc ii} $\lambda$5411 are however clearly {\it asymmetric} toward
redward velocities.
In a further twist, the right-hand panels in 
Fig. 4
show a case where the He{\sc i} $\lambda$5876 line profile transforms
from an absorption profile to a clear P~Cygni profile with redward emission.
In this case the corresponding He{\sc ii} $\lambda$4686 pair exhibit
almost no change in total equivalent width but the peak emission shifts
blueward by $\sim$ 30 km s$^{-1}$. But, once more, the He{\sc ii} $\lambda$5411
profile changes are almost entirely at redward velocities.

To gain further insights into the implications of these
overall profile changes, we examined model predictions using the unified
non-LTE, line-blanketed model atmosphere code CMFGEN (e.g. Hillier
{\&} Miller 1998). Briefly, CMFGEN solves the non-LTE radiation
transfer problem assuming a chemically homogeneous, spherically
symmetric, steady-state outflow. Each model is defined by the stellar
radius, the luminosity, the mass-loss rate, the wind terminal
velocity ($v_\infty$), the stellar mass and by the abundances of the species
included in the calculations. The code does not solve for the hydrodynamical
structure, hence the velocity field has to be defined
using the output of a plane-parallel
model (TLUSTY in this case; see Hubeny {\&} Lanz, 1995)
to define the pseudo-static photosphere,
connected just below the sonic point to a beta-type velocity
law to describe the wind regime.
We calculated
a substantial grid of CMFGEN model atmosphere spectra
and explored sequences of different mass-loss and clumping.
The mass-loss rate has been incremented between
$\sim$ 7 $\times$ 10$^{-9}$ M$_\odot$ yr$^{-1}$
to
$\sim$ 1.6 $\times$ 10$^{-8}$ M$_\odot$ yr$^{-1}$.
We also adjusted the clumping factor, $f_{cl}$,
which measures the over-density inside the clumps
with respect to the average wind density.
Finally, the velocity at which optically thin clumping starts
($v_{cl}$)
near the stellar surface was adjusted between 10 km s$^{-1}$ to
40 km s$^{-1}$, using an exponential law defined by the
parameters $f_{cl}$ and $v_{cl}$.
In each sequence all but one of the above parameters are
kept fixed. The model sequences are shown in Fig. 5 for
He{\sc ii} $\lambda$4686, He{\sc ii} $\lambda$5411 and
He{\sc i} $\lambda$5876.
Comparisons between this grid with variable mass-loss
rates and wind clumping sequences lead us to conclude that the
absorption versus emission strength changes 
seen in the CFHT and ESO data in the left-hand panels in Fig. 4
may qualitatively be explained by (global) mass-loss differences,
but not the asymmetric behaviour evident in He{\sc ii} $\lambda$5411.
Furthermore the relatively large emission line shift seen in the right-hand panels
in Fig. 4 {\it cannot} be explained by line-synthesis model predictions with
a spherically homogeneous wind.

\begin{figure*}
 \includegraphics[scale=0.39]{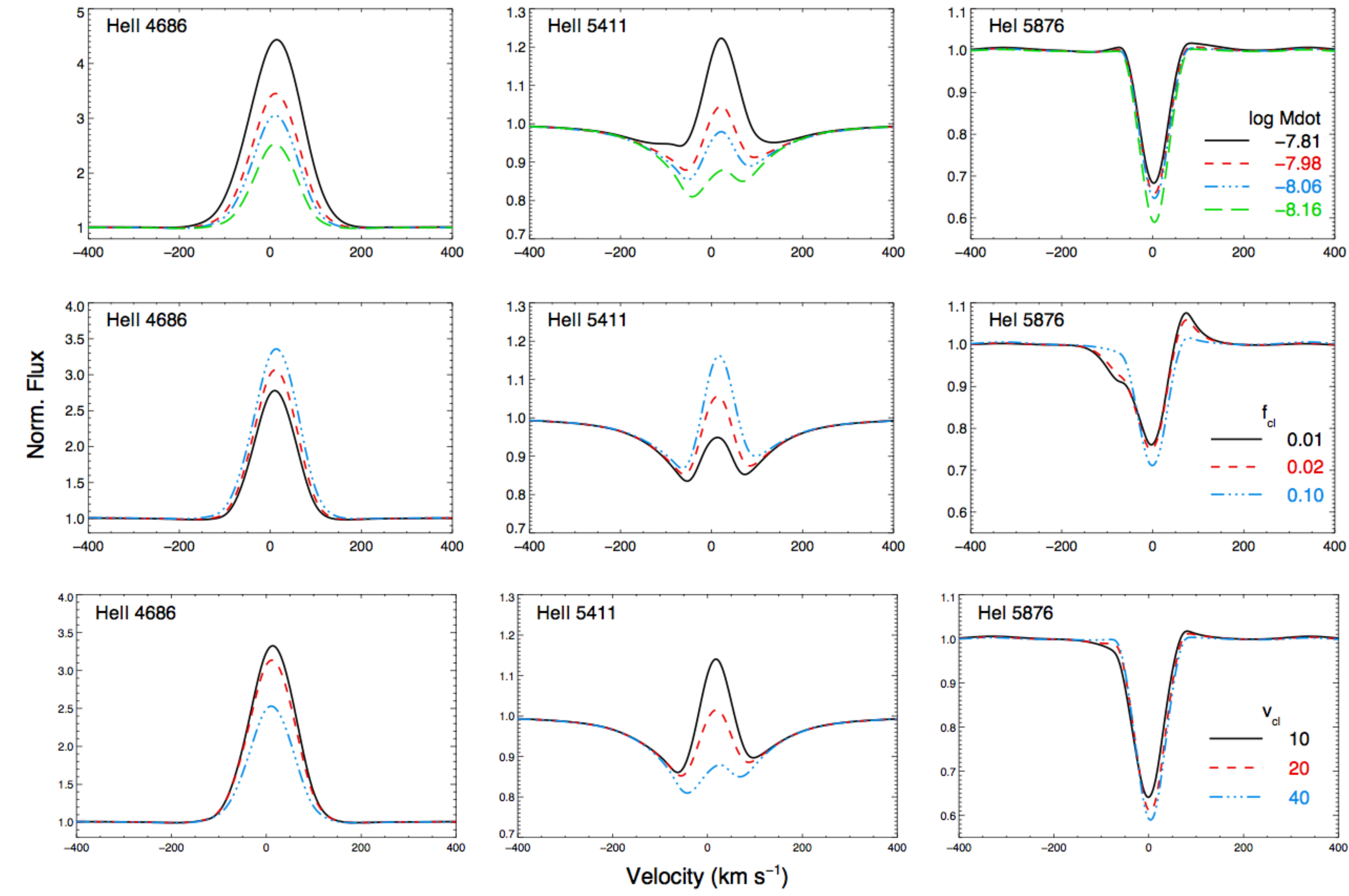}

 \caption{ Sequences of CMFGEN model profiles for
different (i) mass-loss rates (upper row),
clumping factor (middle row)
and velocity
at which clumping starts (bottom row).
Line synthesis profiles are shown for
He{\sc ii} $\lambda$4686 (left-hand column),
He{\sc ii} $\lambda$5411 (middle column) and
He{\sc i} $\lambda$5876 (right-hand column).
}

\end{figure*}

\section{Deep-seated and photospheric changes}

Ordinarily the search for radial velocity shifts in symmetric
photospheric absorption lines would be a relatively straightforward
exercise in hot stars. Trends in velocity changes may then betray
for example the causal role of stellar binarity or multiplicity.
The situation in NGC~2392 is more complex however for two
key reasons;
(i) we have demonstrated in Sect. 3 that the nucleus of NGC~2392
drives a highly variable fast wind, and the origin of these changes
are deep-seated and close to the stellar surface. This means that
a large selection of the He {\sc i}, He {\sc ii} and metal
absorption lines in the optical spectra may be disturbed by
the effects of a variable wind;
(ii) in their analysis of the 3-D and kinematic structure of the
nebula, Garcia-Diaz et al. (2012) present evidence that the inner
nebular shell has an almost pole-on orientation, such that the
inclination angle with respect to the line-of-sight is only
$9\degr$. Assuming that the central star has the same orientation,
the geometry is obviously not favourable for the detection of
binary induced radial velocity motion of the nucleus.

Taking up this challenge nevertheless, we have examined weak, relatively
symmetrical stellar absorption lines in NGC~2392 for evidence of radial
velocity shifts.
The CMFGEN line-synthesis models (Sect. 3) suggest that most of the
optical lines in NGC~2392 are potentially affected by mass-loss and wind clumping
changing. Our grid of models points to the N{\sc iv} $\lambda$6380.8
absorption line as a rare example of a primarily photospheric absorption
line, suitable for a study of subtle radial velocity fluctuations.
Inspection of the time-series data suggests that the N{\sc iv} line 
{\it does}
in fact shift in central velocity by $\sim$ 10 to 15 km s$^{-1}$
in the ESO and CFHT datasets, over hourly timescales.
Pairs of N{\sc iv} profiles separated by $\sim$ 2 hours are
shown in Fig. 6 where the central minimum and blue and red wings are shifted
by $\sim$ 15 km s$^{-1}$.
 A particularly useful consistency check is provided by
H$\epsilon$, since the Ca{\sc ii}\,H $\lambda$3968.5 interstellar/circumsystem
lines provide an excellent nearby fiducial for the accuracy
of the wavelength scale close to H$\epsilon$.
We note in Fig. 6 that the velocity behaviour seen in
N{\sc iv} is systematically mimicked in
H$\epsilon$, thus firming confidence in the notion that there is
a genuine radial velocity shift in NGC~2392.

The N{\sc iv} line profiles were fitted with Gaussian model profiles with least
squares to estimate the central absorption velocities. For a typical internal
fitting error of $\pm$ 1 km s$^{-1}$, the mean and
median absolute deviation (MAD) of the
ESO and CFHT data are 
6.4; MAD $\sim$ 3.7 km s$^{-1}$ and
3.9; MAD $\sim$ 3.4 km s$^{-1}$.
The time-series data are unfortunately not intensive enough over
a sufficiently extensive time-scale to allow for a confident search
for periodic behaviour in the recorded radial velocities.
With this caveat, Fourier power spectra were calculated independently for
the CFHT and HARPS measurements and the results are shown in
Fig. 7 (upper panel).
There are undoubtedly spurious features and aliases throughout the power spectrum,
but it is interesting to note that in both datasets the maximum power occurs
at very similar frequencies i.e. $\sim$
8.1 d$^{-1}$.

\begin{figure}
 \includegraphics[scale=0.23]{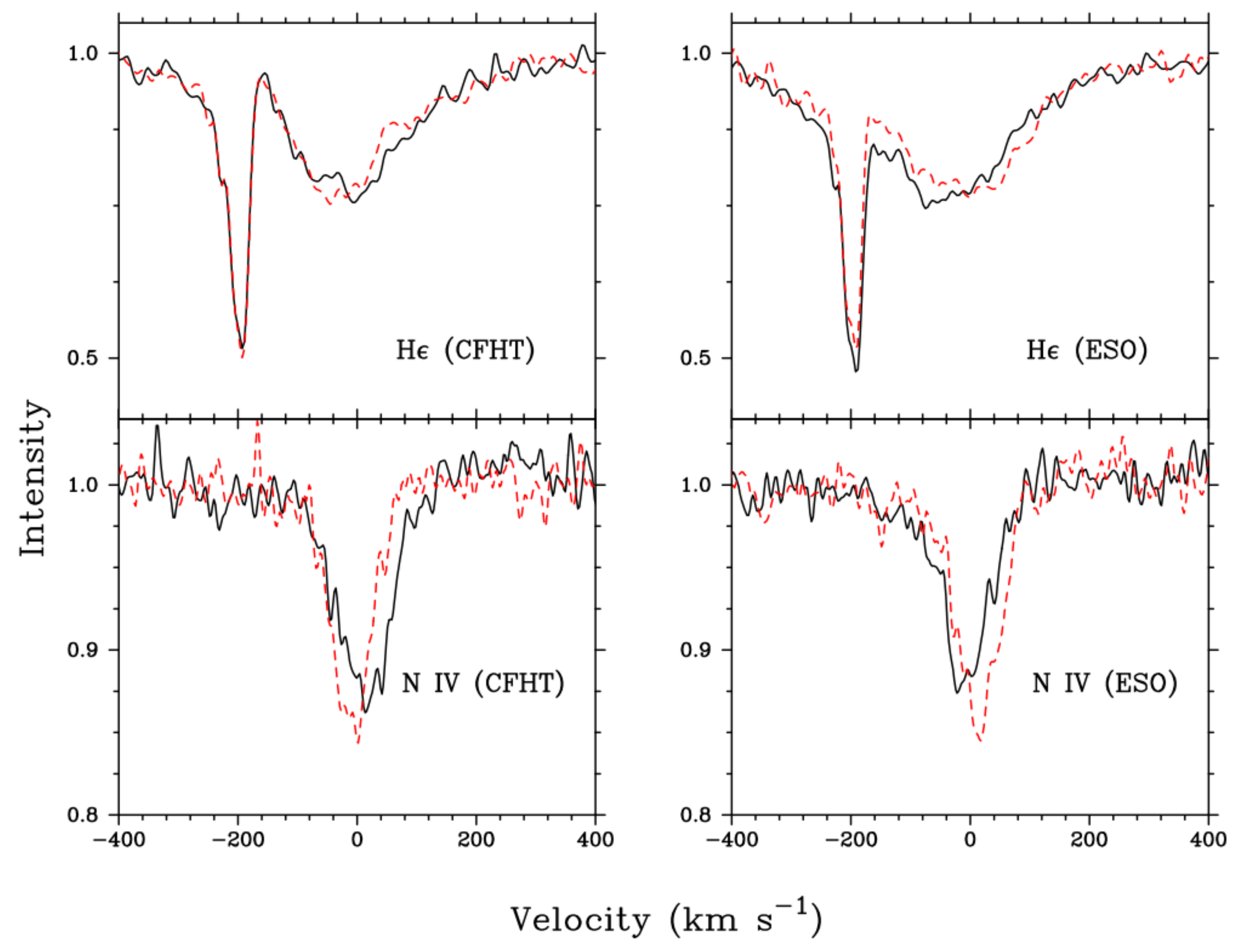}

 \caption{Evidence for radial velocity shifts in pairs of
N{\sc iv} $\lambda$6380.8 line profiles
secured $\sim$ 2 hours apart in CFHT and ESO spectra.
Simultaneous velocity changes are also evident
H$\epsilon$, where the 
sharp interstellar feature of Ca{\sc ii}\,H $\lambda$3968.5 is an
excellent wavelength scale fiducial.
}

\end{figure}

For a slightly more sophisticated treatment we also employed the CLEAN algorithm
(Roberts et al. 1987) which is more suited to the temporal analysis of
unequally spaced finite data samples.
Using a gain of 0.5 and 100 iterations, the spectral window function was
iteratively subtracted in the Fourier domain from the raw power
spectrum shown in the upper panel in Fig. 7.
The resultant `CLEANed' power spectrum is shown 
in the bottom
panel in Fig. 7, where the strongest peak frequency is at
$\sim$ 8.13 d$^{-1}$, {\it for both datasets}.
We estimate an uncertainty of at least
$\sim$ 10{\%}
in this frequency based on the
half-width at half-maximum of the main peak in the window function.
These results provide tentative evidence for a period of $\sim$ 0.123-day
in the N{\sc iv} radial velocity changes, with a semi-amplitude of
$\sim$ 10 km s$^{-1}$.
Figure 8 shows the N{\sc iv} central velocities phased on 0.123-day.
(In each case phase 0 is arbitrarily set
to the timing of the first observation.) The results in Fig. 8 suggest
some coherency on the 0.123-day period {\it independently} for
the ESO (2006) and CFHT (2010) datasets,
though there is relatively larger scatter (in the very small
velocity displacements) at some phases.
Ultimately a more intensive, very high-resolution
time-series dataset of the central star is needed to confirm the
significance of this signal.

\begin{figure}
 \includegraphics[scale=0.23]{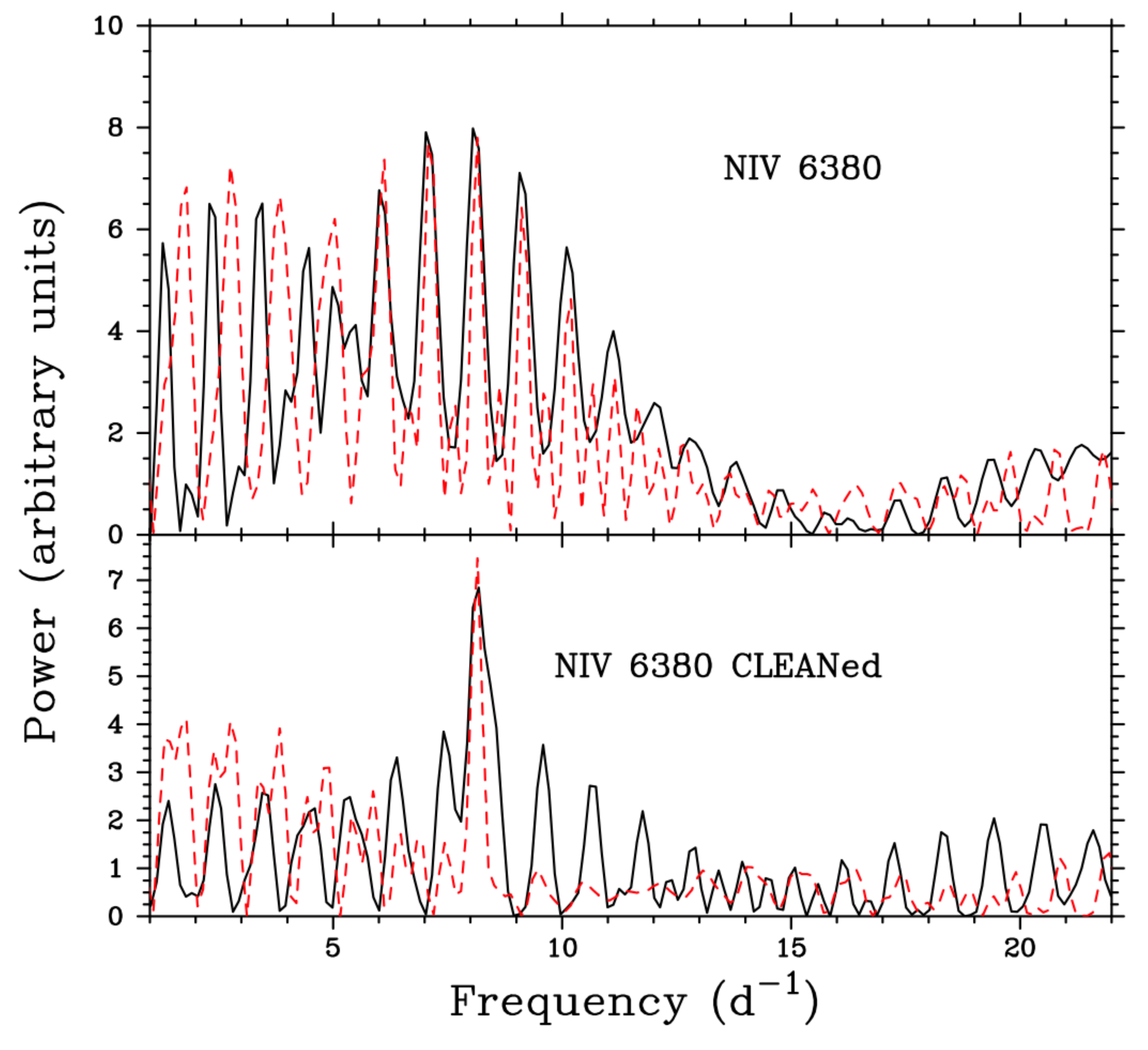}

 \caption{Fourier power spectra for N{\sc iv} $\lambda$6380.8
stellar radial velocity
changes in NGC~2392. The CFHT (solid black) and ESO (dotted red)
power spectra are shown in the upper panels, and corresponding
CLEANed versions are shown in the lower panel. In {\it both} datasets
a peak frequency of $\sim$ 8.1 d$^{-1}$ is apparent. (Power is in
arbitrary units.) 
}

\end{figure}

\begin{figure}
 \includegraphics[scale=0.2]{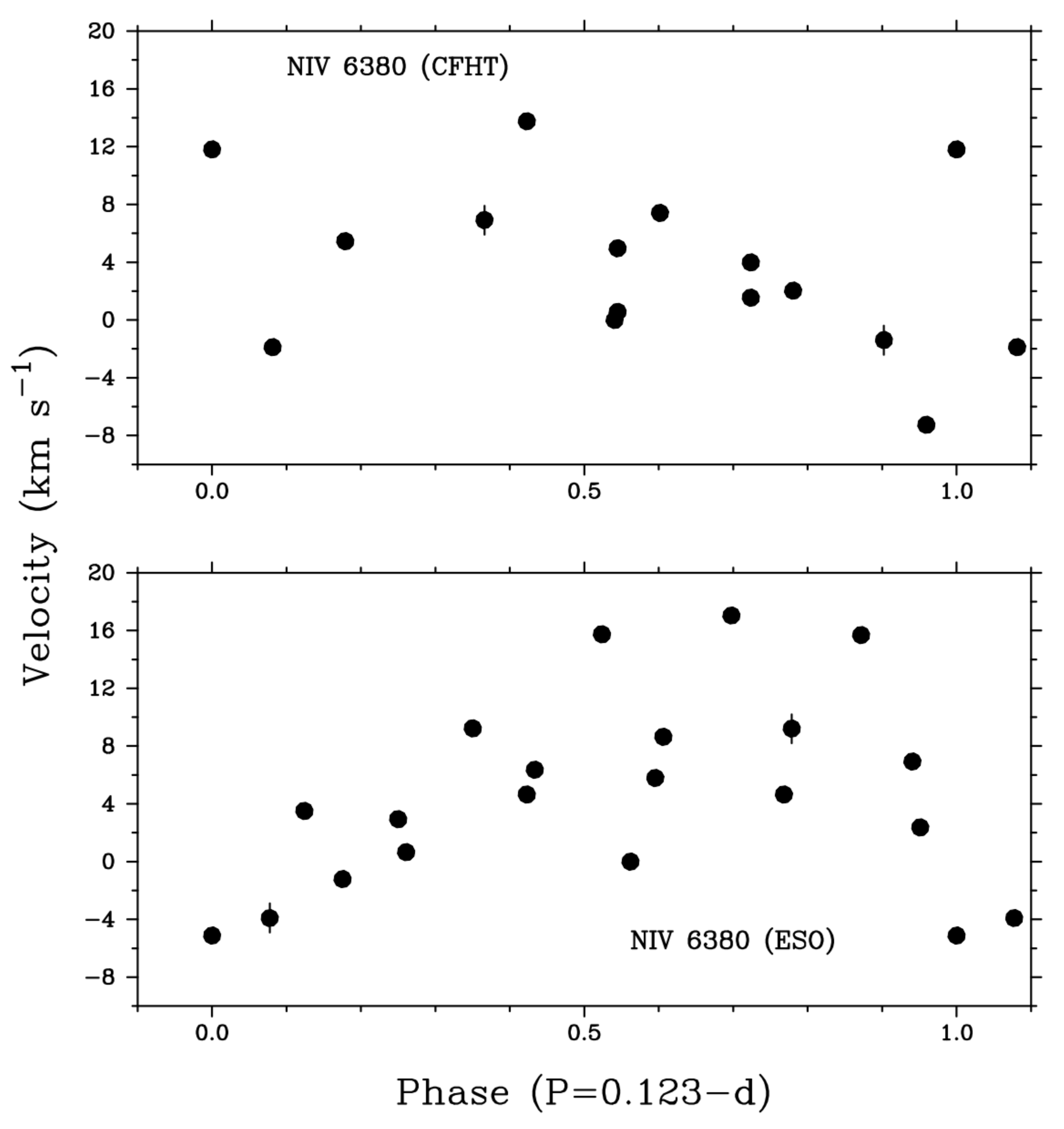}

 \caption{Variations in the central velocity of the
N{\sc iv} $\lambda$6380.8 absorption line phased on the $\sim$ 0.123-d
period.
The ESO HARPS and CFHT ESPaDonS data are separated by $\sim$ 4 years.
Pairs of representative fitting error bars are shown in each panel.
}
\end{figure}

\section{Discussion}
We have provided evidence and demonstrated that in two independently
secured optical spectroscopic time-series separated by
$\sim$ 3 years, both datasets reveal for the central star
of NGC~2392;
(i) stochastic variations in the fast wind-formed recombination
lines on timescales down to $\sim$ 30 min.,
(ii) changes in the overall morphology of the He{\sc i}
and He{\sc ii} line profiles that cannot be accounted for by
1-D line-synthesis predictions for a spherically homogeneous
wind,
(iii) radial velocity shifts of semi-amplitude $\sim$ 10 km s$^{-1}$
in N{\sc iv} $\lambda$6380.8 (and tentatively in H$\epsilon$) that show maximum
Fourier power spectra signal at $\sim$ 0.123-d in ESO (2006) and 
CFHT (2010) data. 

Our detection of radial velocity motion of the central star in NGC~2392
is obviously tentative. Cross-correlation with the majority of stellar
absorption lines in the optical range is not fruitful since most
of features are `contaminated' by the imprints of stellar wind variability
which we have established here.
In advance of more definitive optical spectroscopy and photometry being
secured, we can only speculate on the potential binary components
in NGC~2392: For an assumed circular orbit and
line-of-sight inclination $\sim$ $10\degr$
(e.g. Garcia-Diaz et al. 2012); semi-amplitude of 10 km s$^{-1}$ and
period = 0.123-d (Figs. 8 and 9); assumed (primary) central star mass  
= 0.6 M$_\odot$, the implied dynamical mass of the secondary in 
circular Keplerian
orbit is $\sim$ 0.1 M$_\odot$. Such a low-mass companion would for
example correspond to a late M-dwarf of $T_{\rm eff}$ $\sim$
2000 - 3000K.

\subsection{Constraints from UV line profile morphologies}

The morphologies of the FUV and UV lines are complex in NGC~2392
and provide additional signatures for an asymmetric geometry.
The CMFGEN line-synthesis models (Sect. 3) do not provide
consistent matches to Doppler widths, and absorption and
emission strengths across all the UV ion stages observed in the fast wind.
A selection of UV wind line profiles in NGC~2392 is presented in
Fig. 9, ranging from
O{\sc vi} $\lambda$1031.9 and
S{\sc vi} $\lambda$944.5, to
P{\sc v} $\lambda$1118.0 and
N{\sc v} $\lambda$1238.8, and
C{\sc iv} $\lambda$1548.2 and
Mg{\sc ii} $\lambda$2795.5.
(The data have been retrieved from the {\it FUSE} and {\it IUE}
archives.)
There are some key points to note in Fig 9:
(i) All the wind lines get weaker
with increasing outflow velocity. It may be that the wind plasma
is shifting to a very high ionization state (beyond O{\sc vi})
as it travels to larger radii. Alternatively, the line shapes in Fig. 9
could be an indication that the fast wind of NGC~2392 is moving out
of the line-of-sight, somewhat as may be expected for a polar,
high-latitude wind in an asymmetric geometry;
(ii) The low excitation Mg{\sc ii} line is very narrow
($\sim$ 100 km s$^{-1}$; as is Si{\sc iii} $\lambda$1206.5)
and consistent with a low-velocity equatorial wind
(see e.g. Bjorkman et al. 1994; Massa 1995);
(iii) The presence of P{\sc v} most likely indicates that the
wind is optically very thick, since phosphorous has a low
cosmic abundance and this line would otherwise not be so clearly detected.
However C{\sc iv} and N{\sc v} are weak at intermediate velocities
($\sim$ 200 to 300 km s$^{-1}$). An optically very thick wind that
causes weak absorption can arise in a scenario where the wind is not
covering the entire stellar disk, as may be expected from a polar wind.

We conclude that UV lines provide evidence for an asymmetric, two-component 
outflow in NGC~2392, where 
high-speed high-ionization gas forms preferentially in the polar region.
Slower, low ionization 
material is then confined primarily to a cooler equatorial component of the outflow.

\begin{figure}
 \includegraphics[scale=0.25]{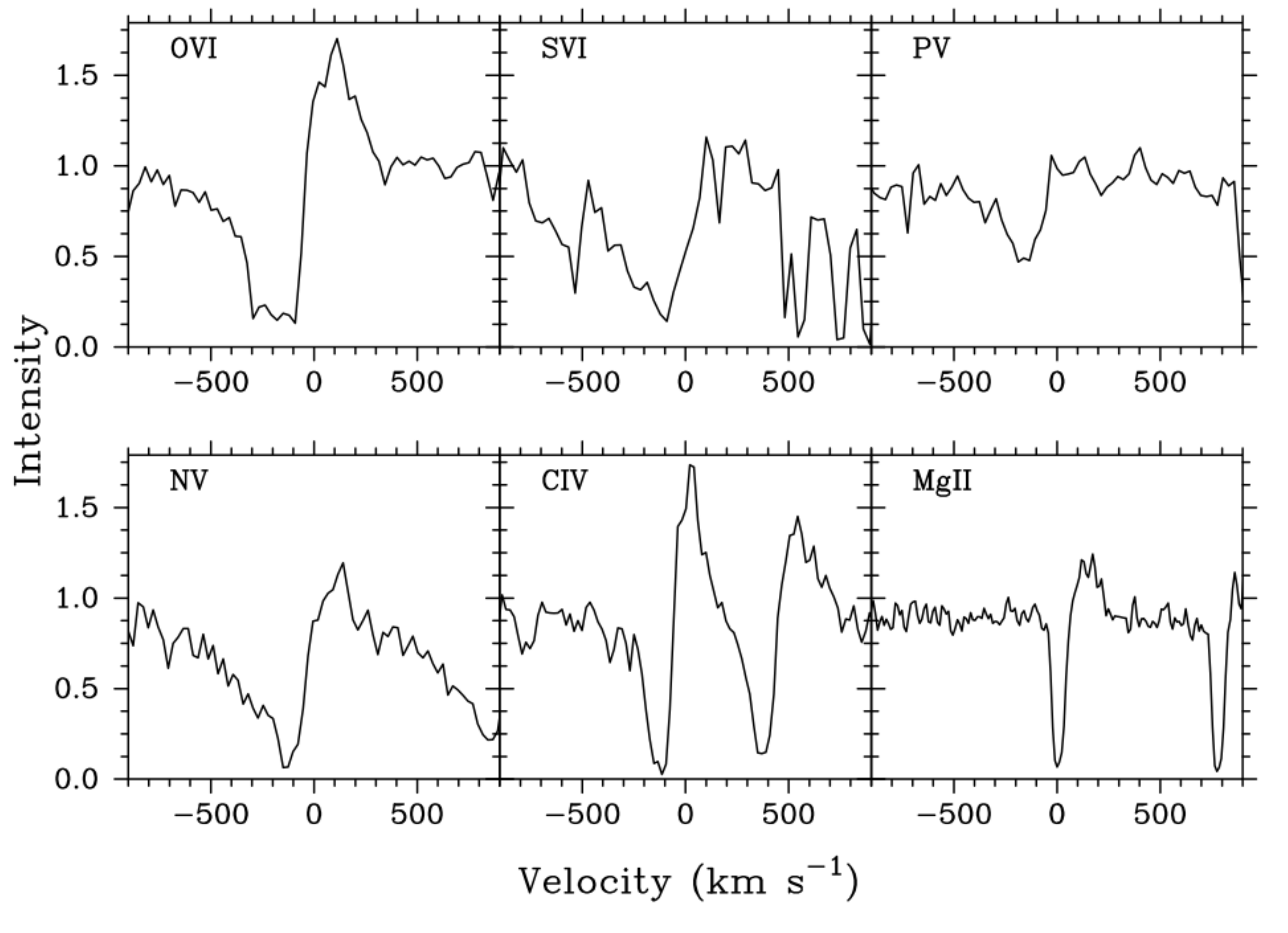}

 \caption{The fast wind in NGC~2392 as revealed by the
FUV and UV resonance line profiles of
O{\sc vi} $\lambda$1031.9,
S{\sc vi} $\lambda$944.5,
P{\sc v} $\lambda$1118.0,
N{\sc v} $\lambda$1238.8,
C{\sc iv} $\lambda$1548.2,
Mg{\sc ii} $\lambda$2795.5.
}

\end{figure}

\section*{Acknowledgments}
Based on observations obtained at the Canada-France-Hawaii Telescope (CFHT) which is
operated by the National Research Council of Canada, the Institut National des Sciences
de l'Univers of the Centre National de la Recherche Scientique of France, and the
University of Hawaii, and 
on observations collected at the European Southern Observatory, 
La Silla (programme ID ESO 076.D-0207(A).
We thank Derck Massa for discussions about the fast wind
of NGC~2392. We acknowledge the helpful comments of the referee.

\bsp

\label{lastpage}

\end{document}